# The Dragon and the Computer: Why Intellectual Property Theft is Compatible with Chinese Cyber-Warfare Doctrine

Paulo Shakarian*, Jana Shakarian, Andrew Ruef

*Abstract:* Along with the USA and Russia, China is often considered one of the leading cyber-powers in the world. In this excerpt, we explore how Chinese military thought, developed in the 1990's, influenced their cyber-operations in the early 2000's. In particular, we examine the ideas of *Unrestricted Warfare* and *Active Offense* and discuss how they can permit for the theft of intellectual property.  We then specifically look at how the case study of Operation Aurora – a cyber-operation directed against many major U.S. technology and defense firms, reflects some of these ideas.

Over the past five years, the news media is seemingly littered with alleged Chinese cyber-incidents. These activities have included instances of theft of guarded scientific data,[1] monitoring of communication of the Dalai Lama,[2] and theft of intellectual property from Google.[3] In a testimony to the Congressional Armed Services Committee, General Keith Alexander, the commander of U.S. Cyber Command and head of the National Security Agency (NSA), stated that China is stealing a "great deal" of military-related intellectual property from the U.S.[4] Clearly, cyber-espionage, which includes the theft of intellectual property, is already a key component of Chinese cyber-strategy.  The recently released report by the security firm *Mandiant* provides technical analysis leading to the conclusion that an organization within the People's Liberation Army (Unit 61398) has been responsible for a great deal of cyber-espionage against English-speaking countries.[5]  In this paper, we highlight some of the relevant Chinese doctrine that we believe led to organizations like Unit 61398 and others.

The activities of exfiltration, monitoring, and theft of digital information described here can be easily labeled as incidents of cyber-espionage. The apparent goal of this type of cyber-operation is not to take the computers offline or destroy the data that they contain but rather to capture data of the opposing force. This being the case, such activities could not be labeled as cyber-attacks, because the targeted systems and their data must remain intact in order to obtain the desired data. Hence, we can define cyber-espionage as the act of obtaining access to data from a computer system without the authorization of that system's owner for intelligence collection purposes.

However, like incidents of computer network attack, these incidents of cyber-espionage too are notoriously difficult to attribute. What then, leads us to believe Chinese involvement in the cyber-espionage incidents? If attribution is so difficult, then why do these actions cause corporations like Google and Northrop Grumman, as well as high-level diplomats such as U.S. Secretary of State Hilary Clinton to issue strong statements against the Chinese government in the wake of such attacks? The issue lies in the origin of the incidents.[6] Often computers involved with the theft of digital information are traced back to networks that are located on the Chinese mainland. Further, forensic analysis of malware from such incidents often indicates the use of Chinese-language software development tools. Though it is virtually impossible to implicate the government of the People's Republic of China (PRC) in these cyber-espionage actions, the fact that they can be consistently traced to the Chinese mainland raises serious policy questions. Is the Chinese government conducting active investigations against the hackers, and what legal actions are they taking once hackers are identified? Is the Chinese government transparently sharing information of these supposed investigations with the victims of the cyber-espionage? What legal actions is Beijing taking to prevent individual hackers from attacking organizations outside of China? These questions must be given serious consideration in the wake of attempted cyber-espionage to when there is evidence of Chinese origin.

What would China have to gain by offering a permissive environment for hackers? It is unlikely that the Chinese government - hallmarked by state monitoring[7] - would not have the resources to reduce such activity. It can further be expected that the fire drawn by the international community is diplomatically undesirable. These activities provide key benefits to the PRC. The nature of the stolen information – which ranges from details of American weaponry and trade secrets to the communications of the Dalai Lama – are all of highly particular interest to Beijing. Further, in the late 1990's and early 2000's several Chinese military thinkers wrote on the topic cyber-warfare.[8] These writings indicate that obtaining unauthorized access to computer systems for the purpose of information exfiltration is an integral part of Chinese cyber-strategy.

To understand Chinese doctrine, we must consider how that nation's culture and traditions have shaped their military thinking in ways vastly different from the West.  In a *SANS* paper,[9] COL Edward Sobiesk highlights an example that illustrates the vast differences in Western and Chinese thought that is noted in a 2002 Report to Congress on *The Military Power of the PRC* by the US Secretary of Defense.[10] This report identifies one of China's strategic objectives as maximizing "strategic configuration of power" called "shi." In the report, a footnote for "shi" states "There is no Western equivalent to the concept of 'shi'. Chinese linguists explain it as 'the alignment of forces,' the 'propensity of things,' or 'potential born of disposition,' that only a skilled strategist can exploit to ensure victory over a superior force." Another interpretation of "shi" could focus on setting favorable conditions. If a nation state attains a higher level of "shi" than a rival, the latter will be easily defeated when conflict does arise, because any battle (if even necessary) will be conducted in conditions extremely favorable to the first nation – as the first nation has already set favorable conditions through the attainment of "shi". By attaining a high-level of access to an adversary's active computer systems – the information stored on those systems has lost two critical aspects – confidentiality and integrity.[11] *Confidentiality* ensures that the information is not viewed by unauthorized individuals, while *integrity* ensures that the information, once retrieved, was not tampered with. Taking away these aspects of an adversary's information can contribute greatly to setting the conditions of the battlefield – perhaps even avoiding battle all-together. Considering "shi" cyber-espionage appears to be a formidable strategic tool – by accessing the opponent's computer systems, the rival's information advantage is reduced while the same is gained on the initiating side.

## *From Active Defense to Active Offense*

Traditionally, the People's Liberation Army (PLA) was focused on the traditional Chinese idea of "active defense" which refers to the idea of not initiating conflict, but being prepared to respond to aggression.[12] In a 2008 Military Review article, Timothy Thomas points out that the late 1990's and early 2000's saw a shift from this mentality, particularly with regard to cyber-warfare. The paradigm that seemed to emerge at this time was "active offense." Under this new rubric, the idea of setting the conditions of the battlefield (i.e. developing "shi") is still pre-eminent but the manner in which it is pursued takes a different turn. In the cyber-arena this entails not only building one's defenses to deter attack, but utilizing cyber-operations to obtain the upper hand in the case of a larger conflict.

This idea of "active offense" is introduced in the 1999 book *Information War* by Zhu Wenguan and Chen Taiyi. In this book, they include a section entitled "Conducting Camouflaged Attacks" where pre-emption and active-offense are laid out.[13] A key component of active-offense is network surveillance which includes obtaining an understanding of an opponent's command and control (C2), electronic warfare (EW), and key weapon systems. In 2002 and 2003, General Dai Qingmin echoes some of these ideas.[14] He stresses that it is necessary for information and cyber operations to be both "precursory" (i.e. done before operations take place) and "whole course" (performed throughout the operation). Where does cyber-espionage fit into this schema? Pre-emption can take many forms. For instance, Russian hackers leveraged denial of service cyber-attacks in the early phases of the Georgia campaign to hamper the opposing force's government, banking, and news media websites. However pre-emption can take other, more subtle forms as well. For example, having constant access to the Tibetian information systems would certainly be an advantage and would perhaps yield the possibility to avoid open conflict altogether. Theft of military secrets relating to new weapon systems may give the Chinese the technical intelligence (TECHINT) needed to find vulnerabilities, or even develop their own copies of said weapons. Stealing intellectual property from software vendors may give Chinese hackers a wealth of insight needed to identify new vulnerabilities for future cyber-attack and cyber-espionage operations.

The work *Information War* and the writings of General Dai illustrate the importance of the cyber-aspect to Chinese military operations. However, many of the cyber-espionage incidents that we will discuss in this paper deal with theft of information from private companies during peace-time. How is this accounted for in the Chinese literature on cyber-warfare? Answers to questions of this type seem to lay in the 1999 book *Unrestricted Warfare* by PLA Colonels Qiao Liang and Wang Xiangsui.[15] In this work, the authors assert that modern warfare extends beyond simply a military domain. Modern warfare includes political, scientific, and economic leaders in addition to military personnel. The notion of "unrestricted" warfare extends not only the domains of war, but also the time at which such actions of war can take place. "Military" operations –that now include information, economic, and psychological aspects, can take place in peacetime in this perspective – further supporting the notion of "active offense." This may help explain why the early 21st century has been littered with stories of Chinese cyber-espionage against corporations and scientific laboratories.

In this same vein, Colonel Wang Wei and Major Yang Zhen of the Nanjing Military Academy's Information Warfare and Command Department wrote in *China Military Science* that in a war against an information-centric society, a nation's political system, economic potential, and strategic objectives will be high-value targets.[16] They then go on to describe that the preferred method to attack such a society would be through the use of asymmetric warfare techniques. Asymmetric warfare refers to the ability of a combatant to defeat a superior force by using tactics that exploit a major weakness in their weapon systems, tactics, or information technology. In the America's war in Iraq from 2003-2011, insurgent often used asymmetric attacks such as road-side bombs as opposed to more traditional attacks that would otherwise expose them to the superior firepower of the Americans. Colonel Wei and Major Zhen espouse asymmetric attacks on a more strategic level – specifically calling for peacetime operations which have military and economic goals. To achieve such goals, under "informatized conditions" they state that both economic and trade warfare must be carried out.[17] Clearly, these authors were influenced by the earlier ideas of *Unrestricted Warfare*. It seems that the peacetime cyber-espionage operations launched from the Chinese mainland against scientific, military, and commercial targets align well with this line of thinking.

Another line of thought in Chinese writing to justify their seemingly bold moves in cyber-space is that they believe these activities can be done with relative impunity. In a 2009 article in *China Military Science,* Senior Colonel Long Fangcheng and Senior Colonel Li Decai state that cyber-operations directed against social, economic, and political targets can be done without fear of such activities leading to large-scale military engagements.[18] As such is the case, they generally regard cyber-warfare as an element of soft-power – albeit one with great effects. They then proceed to claim that the ultimate effect of this highly effective form of soft-power is that the line between peacetime and wartime becomes blurred. This blurring may be a hallmark of cyber-operations in general and might lead to the metaphorical endless war in the near future.

### *INEW and Cyber in the PLA*

The general information warfare (IW) strategy in use by the PLA is known as Integrated Network Electronic Warfare (INEW).[19] This strategy was originally outlined in a book by General Dai Qingmin in 1999 known as *On Information Warfare.* This integration of cyber-operations to traditional information warfare assets is a key component of the INEW strategy. INEW relies on simultaneous application of both electronic warfare and cyber-operations to overwhelm an adversary's command, control, communication, computers, intelligence, surveillance, and reconnaissance (C4ISR). Hence, the mission of key pieces of cyber-warfare (cyber-attack, cyber-espionage, and cyber-defense) – are assigned to elements of the PLA General Staff traditionally given similar roles in electronic warfare.

The General Staff of the PLA is divided into several departments. INEW generally assigns offensive tasks (cyber-attack and more conventional electronic counter measures (ECM)) to the 4th Department – which has traditionally played a large role in offensive information warfare.[20] Notably, General Dai Qingmin was promoted to the head of the 4th Department in 2000 – perhaps an indication that the PLA intended to adopt his vision of INEW. Defensive and intelligence tasks – specifically cyber-defense and cyber-espionage are assigned to the 3rd Department – which traditionally focused on signal intelligence (SIGINT).[21] It is thought that the 3rd Department is the headquarters for the Technical Reconnaissance Bureaus, whose normal mission is SIGINT collection. In the late 1990's several of these Bureaus received awards relating to research in information warfare.[22] Some analysts believe this indicates their role in cyber-operations.[23]

To augment the information warfare specialists in the 3rd and 4th GSD's the Chinese have also established information warfare militia units.[24] These militias can be thought of as a "cyber national guard" as they consist largely of personnel from the commercial information technology (IT) and academia. Open-source reporting indicates that these units have been created from 2003-2008 in Guangzhou, Tianjin, Henan, and Ningxia provinces.[25] There is even evidence that some of these militia received specific wartime tasks – most of which appear to be focused on cyber-attack.[26]

The main ideas of Chinese cyber-operations grew out of the writings of PLA officers in the late 1990's and ultimately implemented in the INEW strategy, which aligns cyber-attack and cyber-espionage responsibilities with organizations conducting similar operations in the realm of electronic warfare.[27] Though the Chinese hacker community came to prominence in the late 1990's and early 2000's with attacks that seemingly had goals congruent to the government, the PRC ultimately disapproved of these actions.[28] As a result, many of the hackers turned "white hat" by either transforming their hacker groups into consulting firms or by obtaining employment with the government and/or academia.[29] Chinese academia also appears to be highly involved with cyber-warfare – not only in research but also potentially with operations.[30]

## *A Case Study in Cyber War through Intellectual Property Theft: Operation Aurora*

On January 12th, 2010, Google announced shocking news. The firm published on its official blog that it had been the victim of a cyber-warfare originating from China. According to the blog, the purpose of the operation was to access the Gmail email-accounts of Chinese human rights activists.[31] As a result of this cyber-espionage operation, Google announced that it would no longer censor results on its flagship search engine in China – google.cn – a move that caused consternation with the PRC. The company stated that if they could not run their search engine uncensored, they would be willing to close operations in China.

Literally minutes after the announcement from Google, Adobe - another major software vendor - announced that their corporate systems had also been hacked.[32] It turns out that both Google and Adobe were targets of the same adversary – an adversary that conducted the very same operation against thirty-two more companies. These firms included Dow Chemical, Northrop Grumman, Symantec, and Yahoo.[33] It seems the purpose of the operation was to exfiltrate not only information about Chinese human rights activists, but intellectual property as well – namely source code of commercially developed software.[34]

This operation – known as "Operation Aurora" - is the topic of this section. It leveraged social engineering along with an advanced Trojan known as Hydraq to steal intellectual property. Several analysts strongly suspect PRC involvement. Here we review the attack, review the evidence of PRC involvement, and discuss the implications of intellectual property theft from corporations.

This act of cyber-espionage employed a vulnerability in Microsoft Internet Explorer that was exploited by software referred to as *Trojan.Hydraq* by the Security firm Symantec. As with several of the cyber-espionage operations discussed in this paper, Operation Aurora was initiated with spear phishing. In the case of the Google break-in, it is thought that this initial spear phishing was directed at an employee using the Microsoft Messenger instant chat software. The user supposedly received a link to a malicious website during one of his chat.[35] It is unknown, if the operations against the other firms were also initiated with chat software. Based on similar operations it seems likely that email may have also been used as a way to initiate the infiltration of the malicious software. Either way, the initial communication to these firms had three characteristics. First, they were sent to a select group of individuals, which suggests that this type of targeting (spear phishing) indicates that the hackers had some additional source of intelligence on their targets. Second, the communications were engineered in a way to appear as though they originated from a trusted source, which also shows that the perpetrators were operating with profiles of their targets. Third, they all contained a link to a website – clicking upon which initiated a certain series of events.

Once the user clicked on the link, their web browser would visit a site based out of Taiwan. This website, in turn, executed malicious JavaScript code – this is source code that runs on a website normally used to provide interactive features to the user. The malicious JavaScript code exploited a weakness in the Microsoft Internet Explorer web browser that was largely unknown at the time. Often such a new vulnerability is termed a "zero-day exploit". The malevolent JavaScript code proceeds to download a second piece of malware from Taiwan - disguised as an image file. This secondary malicious software would proceed to run in Windows and set up a *back door* allowing a cyber-spy access to the targeted system.[36] A back door refers to a method of accessing a system that allows an intruder to circumvent the normal security mechanism. The use of a zero-day exploit is significant because identifying such a vulnerability most likely required a skillful engineering effort. This, along with the highly-targeted spear-phishing campaign (suggesting that the hackers had access to some additional intelligence on their targets), might hint at the backing of a larger organization – possibly a nation-state.

### *Theft of Intellectual Property*

Several months after Google announced that it had been hacked, the New York Times reported that more than just email accounts of Chinese human rights activists had been compromised. Citing an unnamed source with direct knowledge of the Google investigation, reporter John Markoff wrote that the source code to Google's state of the art password system had likely been stolen during Operation Aurora.[37] The system, known as *Gaia*, was designed to allow users of Google's software to use a single username and password to access the myriad of Google services. This software is also known as "Single Sign-On." Markoff reported that Google addressed the problem by adding an additional layer of encryption to their password system.

The compromise of *Gaia* is significant for more than one reason. First, obtaining software source code of a commercial system is intellectual property theft and thus unlawful in the U.S. As with the data stolen during Titan Rain, the stolen source code could allow certain developers to illicitly create software similar to *Gaia*. If we view Operation Aurora as the actions of a nation state, theft of intellectual property can be thought of as a form of economic warfare – leveling the technological playing field in order to reduce the advantage of an adversary nation's industrial capability. Clearly, this is in line with the Chinese ideas of *Unrestricted Warfare* – where various forms of information warfare occur constantly (including during peacetime) and attack all aspects of a nation's power (including industry).

However, beyond the economic advantages gained by the theft of source code, major security implications are also imminent – particularly in the case of *Gaia*. For instance, analysts working with the hackers would most likely determine technical vulnerabilities in the password system.

Though it is clear that the theft of intellectual property is an important consideration for corporations, it also raises an important question. How were the attackers able to obtain source code for a system such as *Gaia* by leveraging a relatively small number of compromised computer systems? It turns out that many corporations work with specialized servers as store-houses for this type of data – often fittingly termed "intellectual property repositories." Centralized locations of this type of data make it easier for teams to work collaboratively on a project and share information with each other. These systems often take the form of Software Configuration Management (SCM) systems such as IBM Rationale© or content management systems such as Microsoft SharePoint©.

Operation Aurora invalidated a key assumption made by many system administrators and IP repository software vendors at the time. The professionals operating those networks assumed that the intellectual property would not be accessed due to security countermeasures taken to protect the network as a whole. The result of this perspective is a lesser focus on the security of an IP repository lying within the perimeter of a corporation's network. By utilizing a zero-day vulnerability for their mission, the perpetrators behind Operation Aurora were able to exploit this assumption.

Theft of intellectual property presents another key difficulty – determining what was actually stolen. In the wake of Operation Aurora, security researcher George Kurtz wrote an article entitled "Where's the body?"[38] As opposed to a physical theft where it is relatively easy to determine what was stolen, with cyber-espionage and data-exfiltration this is much more difficult to establish. Though systems administrators have a few tools at hand - such as the examination of server logs and the analysis of network traffic - in advanced cyber-espionage operations hackers often take various steps to cover-up their tracks and operate in a manner, which makes it difficult to ascertain what data was stolen. Though security vendors provide software solutions to help with this issue, determining "where's the body" in the wake of a cyber-espionage operation is still often a difficult task.

### *Indicators of PRC Involvement*

It is interesting that Google's announcement of the security breach seems to implicate Chinese involvement – or suggests at least complacency on the side of the government. Here are some indicators that Operation Aurora was executed with the full knowledge or even under the directive of the Chinese government.

The earliest signs of Chinese involvement were made public in January 2010 – several weeks after Google's initial blog post. A report released by the security firm VeriSign stated that the "source IP's and drop server of the attack correspond to a single foreign entity consisting of either agents of the Chinese state or proxies thereof."[39] The researchers at VeriSign also found that the Aurora hackers used HomeLinux DynamicDNS and "borrowed" IP addresses from the American firm Linode (a company specializing in Virtual Private Server Hosting). These are the same circumstances as in a July 2009 DDoS attacks against South Korea and Washington, D.C. When considered with other similarities, VeriSign researchers concluded that Aurora and the attacks against Washington, D.C. and South Korea were possibly conducted by the same entity.

Just a few weeks later, New York Times reporters John Markoff and David Barboza published an article which stated that investigators had identified two Chinese schools of higher education involved in the attack[40] - Shanghai Jiaotong University and the Lanxiang Vocational School. The former's Information Security Engineering Institute is the workplace of Peng Yinan (alleged to be the Chinese hacker "CoolSwallow"). When New York Times reporters conducted an anonymous telephone interview with a professor from that institute, they were surprised with the candid response. He stated that students hacking into foreign computer networks were "quite normal".[41] However, as an alternate explanation, the professor stated that the university's IP address could also have been hijacked which he said "frequently happens".[42]
At the Lanxiang Vocational School, the investigators were able to identify a specific class taught by a Ukrainian professor suspected to be involved in Operation Aurora.[43] When confronted with the suspicion, the dean of the computer science department there (identified in the media only as Mr. Shao) stated that the students at the school simply would not have the ability to carry out such an attack. However, he did acknowledge that students from the school were often recruited into the military.[44]

The reports of Chinese involvement might have inspired U.S. Secretary of State Hillary Clinton's speech on Internet freedom given shortly after Google's announcement.[45] In this speech she called upon China to perform a transparent investigation on the intrusions into Google. This was perhaps the strongest statement by a high-ranking U.S. government official made in response to a cyber-warfare incident at the time.

Operation Aurora illustrates the continued evolution of cyber-espionage in the early 21$^{st}$ century. In this case of cyber-espionage the targeted information was deemed so important that the operators utilized a zero-day exploit and spear-phishing to gain access to corporate systems, locate the target's intellectual property repositories, and steal company secrets. Originally reported by Google, this operation affected over thirty big-name companies. The stolen information was unlikely to only further economic gain, but is also feasibly beneficial for technical intelligence, such as the evaluation of vulnerabilities – possibly for use in further cyber-attacks. Operation Aurora invalidated existing assumptions about the security of intellectual property repositories in corporations and again highlighted the difficulty of determining the specifics of the captured data. The news media accounts of potential involvement of China led to a diplomatic statement by the U.S. Secretary of State. Operation Aurora is not unique. In its aftermath, there have been other Chinese-attributed cyber maneuvers performed with the goal of stealing intellectual property. A series of events known as Nitro[46] (directed against the chemical industry) and Night Dragon[47] (against the energy sector) are but two examples. Finally, there are many potential second and third order effects of a major software vendor such as Google or Adobe being hacked. It is unknown what consequences the knowledge of potentially widely-used software, such as Google's *Gaia* password system, will have in follow-on cyber-operations. Though currently not connected to Aurora, it was recently revealed that Adobe's software certificate system was hacked – allowing malicious software to create seemingly safe add-ons too many of that firm's software.[48] In this case, a development server at Adobe was broken into. It is a clear example of how the cyber-security of a major software vendor's own systems can have a direct impact on an extremely large population of users – hence potentially providing ample opportunities to an adversary conducting follow-on cyber-attacks.

----

Here we discussed several ideas espoused by China's military thinkers on information warfare – highlighting the ideas of *Unrestricted Warfare* – in which cyber operations are thought to extend into peacetime and involves military, political, economic, and scientific domains. We looked at how the Chinese structured their cyber-warriors around the INEW strategy. In the PLA, cyber-operations were put under the responsibility of organizations with similar missions in the realm of electronic warfare. Finally, we saw how some of these ideas may have been put into practice with Operation Aurora where a zero-day exploit allowed operators to steal intellectual property from repositories at Google, Adobe, and many other major companies in late 2009.

## Notes

[1] Steve DeWeese, Bryan Krekel, George Bakos, Christopher Barnett, Capability of the People's Republic of China to Conduct of Cyber Warfare and Computer Network Exploitation, Northrop Grumman, October 2009.

[2] Information Warfare Monitor, *Tracking Gh0stNet: Investigating a Cyber Espionage Network*, March, 2009.

[3] Kim Zetter, "Google Hackers Targeted Source Code of More Than 30 Companies," Wired Threat Level, 13 Jan. 2010, accessed 8 Jan. 2012. Available at: http://www.wired.com/threatlevel/2010/01/google-hack-attack/.

[4] J. Nicholas Hoover.(2012, March, 27).NSA Chief: China behind RSA Attacks.*InformationWeek. Retrieved from:* http://www.informationweek.com/news/government/security/232700341?cid=RSSfeed_IWK_All.

[5] "APT1: Exposing One of China's Cyber Espionage Units," *Mandiant.* Retrieved from http://intelreport.mandiant.com/.

[6] Origin cannot only refer to the source-IP addressed (traced through intermediate proxies) but also the origin of the software as determined by technical analysis of the code (i.e. the origin based on the version of the compiler and the language of the operating system used to create the software in question).

[7] "Chinese Internet Giants Agree to Help Government Monitor Information," *Voice of America News,* 5 Nov., 2011, http://www.voanews.com/content/chinese-internet-giants-agree-to-help-government-monitor-information-133327903/168182.html (accessed 25 Mar. 2013).

[8] Timothy Thomas, "China's Electronic Long-Range Reconnaissance," *Military Review,* November-December 2008, 47-54.

[9] Edward Sobiesk, "Redefining the Role of Information Warfare in Chinese Strategy" *SANS Institute InfoSec Reading Room,* March 2003, http://www.sans.org/reading_room/whitepapers/warfare/redefining-role-information-warfare-chinese-strategy_896 (accessed 22 December, 2011).

[10] Office of the Secretary of Defense of the United States of America, Report to Congress on *The Military Power of the People's Republic of China,* 12 July 2002, 5-6.

[11] W. V. Maconachy, Corey D. Schou, Daniel Ragsdale, Don Welch, "A Model for Information Assurance: An Integrated Approach," *Proceedings of the 2001 IEEE Workshop on Information Assurance and Security*, June 2001, http://it210web.groups.et.byu.net/lectures/MSRW%20Paper.pdf (accessed 22 December, 2011).

[12] Timothy Thomas, "China's Electronic Long-Range Reconnaissance," *Military Review,* November-December 2008, 47-54.

[13] Ibid.

[14] Ibid.

[15] Sobiesk, 8.

[16] Timothy Thomas, "Google Confronts China's Three Warfares," *Parameters,* Summer 2010, 101-105.

[17] Wang Wei and Yang Zhen, "Recent Development in the Study of the Thought of People's War under Informatized Conditions," China Military Science, 2d iss. 2009.

[18] Long Fangcheng and Li Decai, "On the Relationship of Military Soft Power to Comprehensive National Power and State Soft Power," China Military Science, Issue 5, 2009, 120-29.

[19] Steve DeWeese, Bryan Krekel, George Bakos, Christopher Barnett, *Capability of the People's Republic of China to Conduct of Cyber Warfare and Computer Network Exploitation*, Northrop Grumman, October 2009.

[20] Ibid.

[21] Ibid.

[22] Ibid.

[23] Ibid.

[24] Ibid.

[25] Ibid.

[26] Ibid.